\documentclass[conference,compsoc]{IEEEtran}

\renewcommand{\IEEEbibitemsep}{2pt plus 0.5pt}
\makeatletter
\IEEEtriggercmd{\reset@font\footnotesize} %
\makeatother
\IEEEtriggeratref{1}

\usepackage{xspace}

\usepackage{cite}

\usepackage{multicol}
\usepackage{multirow}

\usepackage{booktabs}

\usepackage{xcolor}

\usepackage{pifont}

\usepackage[colorlinks=true,urlcolor=blue,citecolor=black]{hyperref}

\usepackage[utf8]{inputenc}
\usepackage[T1]{fontenc}

\usepackage{algorithm}
\usepackage{algpseudocode}

\usepackage{amsmath}

\usepackage[framemethod=TikZ]{mdframed}

\usepackage{paralist}

\usepackage{import}
\usepackage{transparent}

\usepackage{graphicx}
\usepackage{afterpage}

\usepackage{markdown}

\definecolor{sangria}{rgb}{0.57, 0.0, 0.04}
\definecolor{pinegreen}{rgb}{0.0, 0.47, 0.44}
\definecolor{rossocorsa}{rgb}{0.83, 0.0, 0.0}
\definecolor{ao}{rgb}{0.0, 0.0, 1.0}
\definecolor{deepjunglegreen}{rgb}{0.0, 0.29, 0.29}
\definecolor{dartmouthgreen}{rgb}{0.05, 0.5, 0.06}

\definecolor{rubcolor}{HTML}{093158}
\definecolor{amber}{rgb}{1.0, 0.75, 0.0}

\definecolor{arsenic}{rgb}{0.23, 0.27, 0.29}

\global\mdfdefinestyle{insightstyle}{%
roundcorner=3pt
outerlinewidth=0pt,
innerlinewidth=0pt,
backgroundcolor=arsenic!10,
outerlinecolor=black,
linecolor=black,
nobreak=false,
}

\newcommand{\numaeaccesstoreproduced}{57\xspace}

\newcommand{\numaeaccessperconly}{60\%\xspace}

\newcommand{\numcodeavailablewperc}{74\% (214)\xspace}
\newcommand{\numcodeavailableperconly}{74\%\xspace}

\newcommand{\numcodeunavailablewperc}{23\% (66)\xspace}

\newcommand{\numdataavailablewperc}{11\% (31)\xspace}

\newcommand{\numdataavailableinsteadofcode}{20\xspace}

\newcommand{\numaenobadgewperc}{37\% (107)\xspace}

\newcommand{\numaeunavailablewperc}{36\% (103)\xspace}

\newcommand{\numaeanybadgewperc}{23\% (66)\xspace}
\newcommand{\numaeanybadgeperconly}{23\%\xspace}
\newcommand{\numaeavailable}{64\xspace}

\newcommand{\numaefunctional}{63\xspace}

\newcommand{\numaereproduced}{16\xspace}

\newcommand{\numcoverageexplanationyeswperc}{32\% (48)\xspace}

\newcommand{\numdatasetusednonewperc}{61\% (91)\xspace}

\newcommand{\numdatasetusedlavamwperc}{17\% (26)\xspace}

\newcommand{\numdatasetusedfuzzbenchwperc}{10\% (15)\xspace}

\newcommand{\numdatasetusedftswperc}{8\% (12)\xspace}

\newcommand{\numdatasetusedcgcwperc}{5\% (8)\xspace}

\newcommand{\numdatasetusedmagmawperc}{4\% (6)\xspace}

\newcommand{\numdatasetusedunifuzzwperc}{1\% (2)\xspace}

\newcommand{\numavgtargets}{8.9\xspace}

\newcommand{\numtargetsmaxonceusedwperc}{76\% (576)\xspace}

\newcommand{\numuniquetargets}{753\xspace}

\newcommand{\numcompetitoraflwperc}{35\% (53)\xspace}

\newcommand{\numcompetitorqsymwperc}{15\% (23)\xspace}

\newcommand{\numcompetitoraflfastwperc}{14\% (21)\xspace}

\newcommand{\numcompetitorangorawperc}{13\% (20)\xspace}

\newcommand{\numcompetitoraflppwperc}{9\% (14)\xspace}

\newcommand{\numcompetitorfairfuzzwperc}{8\% (12)\xspace}

\newcommand{\nummissingcompetitorsmissingtoolswperc}{20\% (30)\xspace}

\newcommand{\nummissingcompetitorsdoesnotcompareagainstbaselinewperc}{3\% (4)\xspace}

\newcommand{\numbasedonnonewperc}{33\% (49)\xspace}
\newcommand{\numbasedonnoneperconly}{33\%\xspace}

\newcommand{\numbasedonaflwperc}{30\% (45)\xspace}

\newcommand{\numbasedonaflppwperc}{6\% (9)\xspace}

\newcommand{\numbasedonlibfuzzerwperc}{5\% (7)\xspace}

\newcommand{\numbasedonsyzkallerwperc}{4\% (6)\xspace}

\newcommand{\numavgcompetitors}{3.2\xspace}

\newcommand{\numpapersmultipleruntimes}{26\xspace}

\newcommand{\numruntimetwentyfourhourswperc}{56\% (84)\xspace}

\newcommand{\numruntimelessthantwentythreewperc}{27\% (40)\xspace}
\newcommand{\numruntimelessthantwentythreeperconly}{27\%\xspace}

\newcommand{\numruntimenonewperc}{5\% (8)\xspace}

\newcommand{\numruntimemorethantwentyfourwperc}{29\% (44)\xspace}

\newcommand{\numresourceallocationfairwperc}{74\% (111)\xspace}

\newcommand{\numresourceallocationunfair}{8\xspace}
\newcommand{\numresourceallocationunfairwperc}{5\% (8)\xspace}
\newcommand{\numresourceallocationunfairperconly}{5\%\xspace}

\newcommand{\numresourceallocationnotapplicablewperc}{5\% (8)\xspace}

\newcommand{\numresourceallocationnotspecifiedwperc}{15\% (23)\xspace}

\newcommand{\numcoresnonewperc}{25\% (38)\xspace}

\newcommand{\numcoresonewperc}{27\% (40)\xspace}

\newcommand{\numcorestwowperc}{8\% (12)\xspace}

\newcommand{\numuninformedseedsuninformedseedswperc}{25\% (38)\xspace}

\newcommand{\numuninformedseedsnotspecifiedwperc}{25\% (37)\xspace}

\newcommand{\numuninformedseedsinformedseedswperc}{20\% (30)\xspace}

\newcommand{\numuninformedseedsbenchmarkseedswperc}{16\% (24)\xspace}

\newcommand{\numuninformedseedsnoseedsrequiredwperc}{11\% (16)\xspace}

\newcommand{\numuninformedseedsmultipleseedsetswperc}{3\% (5)\xspace}

\newcommand{\numseedfairnesssameseedsforallwperc}{46\% (69)\xspace}

\newcommand{\numseedfairnessnotspecifiedwperc}{30\% (45)\xspace}

\newcommand{\numseedfairnessdifferentseedswperc}{5\% (8)\xspace}
\newcommand{\numseedfairnessdifferentseedsperconly}{5\%\xspace}

\newcommand{\numseedexplanationnoinfoavailablewperc}{50\% (75)\xspace}

\newcommand{\numseedexplanationinfoonseedsavailablewperc}{39\% (59)\xspace}

\newcommand{\numevalmetricscodecoveragewperc}{77\% (115)\xspace}

\newcommand{\numevalmetricsttewperc}{13\% (20)\xspace}

\newcommand{\numevalmetricsbugswperc}{71\% (107)\xspace}

\newcommand{\numcoveragebranchcoveragewperc}{19\% (29)\xspace}

\newcommand{\numcoverageedgeswperc}{17\% (25)\xspace}

\newcommand{\numcoveragebasicblockswperc}{13\% (19)\xspace}

\newcommand{\numcoveragepathswperc}{11\% (17)\xspace}

\newcommand{\numcoveragelinecoveragewperc}{5\% (8)\xspace}

\newcommand{\numcoverageexplanationnodefinitionexplanationwperc}{45\% (67)\xspace}

\newcommand{\nummwunostatisticaltestwperc}{63\% (94)\xspace}
\newcommand{\nummwunostatisticaltestperconly}{63\%\xspace}

\newcommand{\nummwumannwhitneyuwperc}{37\% (55)\xspace}

\newcommand{\nummwufewtrialsfiveorlesswperc}{15\% (22)\xspace}
\newcommand{\nummwufewtrialsfiveorlessperconly}{15\%\xspace}

\newcommand{\numeffectsizevarghaanddelaneystestwperc}{10\% (15)\xspace}

\newcommand{\numeffectsizeotherwperc}{2\% (3)\xspace}

\newcommand{\numeffectsizenonewperc}{88\% (132)\xspace}

\newcommand{\numpapersmultipletrials}{8\xspace}
\newcommand{\numtrialsnone}{21\xspace}

\newcommand{\numtrialslessthantenwperc}{55\% (83)\xspace}

\newcommand{\numuncertaintynomeasureofuncertaintywperc}{73\% (109)\xspace}
\newcommand{\numuncertaintynomeasureofuncertaintyperconly}{73\%\xspace}

\newcommand{\numcvesanalyzed}{339\xspace}
\newcommand{\numcvesanalyzedperconly}{51\%\xspace}
\newcommand{\numpaperscveanalysis}{35\xspace}
\newcommand{\numallcves}{662\xspace}
\newcommand{\numallpaperswithcves}{59\xspace}
\newcommand{\numavgcvesclaimed}{9.7\xspace}

\newcommand{\numcvecategoryfixedwperc}{43\% (145)\xspace}
\newcommand{\numcvecategoryfixedperconly}{43\%\xspace}

\newcommand{\numcvefixedfixedwperc}{42\% (143)\xspace}

\newcommand{\numcvecategoryblindedwperc}{26\% (88)\xspace}
\newcommand{\numcvecategoryblindedperconly}{26\%\xspace}

\newcommand{\numcvecategoryignoredwperc}{20\% (69)\xspace}
\newcommand{\numcvecategoryignoredperconly}{20\%\xspace}
\newcommand{\numcveignoredprojectdead}{14\xspace}
\newcommand{\numcvecategoryinvalidwperc}{11\% (37)\xspace}
\newcommand{\numcvecategoryinvalidperconly}{11\%\xspace}

\newcommand{\numpapersallgoodcves}{18\xspace}
\newcommand{\numpappersallgoodcvescvesum}{67\xspace}

\newcommand{\eg}{e.\,g.,\xspace}
\newcommand{\ie}{i.\,e.,\xspace}
\newcommand{\etal}{et~al.\@\xspace}

\newcommand{\totalpapers}{289\xspace}
\newcommand{\numpapers}{150\xspace}
\newcommand{\numpaperswperc}{52\% (150)\xspace}
\newcommand{\numcasestudies}{eight\xspace}

\newcommand{\cgc}{CGC\xspace}
\newcommand{\lavam}{LAVA-M\xspace}
\newcommand{\magma}{Magma\xspace}

\newcommand{\unibench}{Unibench\xspace}
\newcommand{\fuzzbench}{FuzzBench\xspace}
\newcommand{\fts}{Google's Fuzzer-Test-Suite\xspace}

\newcommand{\revbugbench}{RevBugBench\xspace}

\newcommand{\profuzzbench}{ProFuzzBench\xspace}

\newcommand{\firmafl}{Firm-AFL\xspace}
\newcommand{\afl}{AFL\xspace}

\newcommand{\numthreatstovaliditywperc}{20\% (30)\xspace}

\newcommand*{\inlineincludegraphics}[1]{%
    \raisebox{.2\baselineskip}{%
        \includegraphics[
        height=\baselineskip,
        width=\baselineskip,
        keepaspectratio,
        ]{#1}%
    }%
}

\begin{document}

\title{SoK: Prudent Evaluation Practices for Fuzzing\\ \emph{{\small To appear at IEEE S\&P 2024. Author version.}}}

\author{%
\small
{\rm Moritz Schloegel$^1$, Nils Bars$^1$, Nico Schiller$^1$, Lukas Bernhard$^1$, Tobias Scharnowski$^1$} \\%
{\rm Addison Crump$^1$, Arash Ale-Ebrahim$^1$, Nicolai Bissantz$^2$, Marius Muench$^3$, Thorsten Holz$^1$}\vspace{1.2em}\\%
$^1$\textit{CISPA Helmholtz Center for Information Security, \{first.lastname\}@cispa.de}\\
$^2$\textit{Ruhr University Bochum, nicolai.bissantz@ruhr-uni-bochum.de}\\
$^3$\textit{University of Birmingham, m.muench@bham.ac.uk}
} %

\maketitle

\begin{abstract}
Fuzzing has proven to be a highly effective approach to uncover software bugs over the past decade. After \afl popularized the groundbreaking concept of lightweight coverage feedback, the field of fuzzing has seen a vast amount of scientific work proposing new techniques, improving methodological aspects of existing strategies, or porting existing methods to new domains. All such work must demonstrate its merit by showing its applicability to a problem, measuring its performance, and often showing its superiority over existing works in a thorough, empirical evaluation. Yet, fuzzing is highly sensitive to its target, environment, and circumstances, \eg randomness in the testing process. After all, relying on randomness is one of the core principles of fuzzing, governing many aspects of a fuzzer's behavior. Combined with the often highly difficult to control environment, the \emph{reproducibility} of experiments is a crucial concern and requires a prudent evaluation setup. 
To address these threats to validity, several works, most notably \emph{Evaluating Fuzz Testing} by Klees~\etal, have outlined how a carefully designed evaluation setup should be implemented, but it remains unknown to what extent their recommendations have been adopted in practice.

In this work, we systematically analyze the evaluation of \numpapers fuzzing papers published at the top venues between 2018 and 2023. We study how existing guidelines are implemented and observe potential shortcomings and pitfalls. We find a surprising disregard of the existing guidelines regarding statistical tests and systematic errors in fuzzing evaluations. For example, when investigating reported bugs, we find that the search for vulnerabilities in real-world software leads to authors requesting and receiving CVEs of questionable quality.
Extending our literature analysis to the practical domain, we attempt to reproduce claims of \numcasestudies fuzzing papers. 
These case studies allow us to assess the practical reproducibility of fuzzing research and identify archetypal pitfalls in the evaluation design.
Unfortunately, our reproduced results reveal several deficiencies in the studied papers, and we are unable to fully support and reproduce the respective claims. To help the field of fuzzing move toward a scientifically reproducible evaluation strategy, we propose updated guidelines for conducting a fuzzing evaluation that future work should follow.

\end{abstract}

\section{Introduction}%
\label{sec:introduction}

\emph{Fuzzing}, a portmanteau of ``fuzz testing'', has gained much attention in recent years, and the method has proven to be highly successful in uncovering many types of faults in software systems. 
Companies such as Meta, Google, and Oracle have invested significant resources in this technology and use it to test their products. 
Large software projects such as web browsers or the Linux kernel incorporate fuzzing into their development cycle, and Google is running an extensive and continuous fuzzing campaign for more than $1,200$ open-source projects via OSS-Fuzz~\cite{ossfuzz}.
Beyond the wide acceptance in the industry, a large number of academic papers have proposed numerous improvements and novel techniques to enhance fuzzing further.
More specifically, we found that, over the past six years, more than 280 papers on fuzzing have been published in the top computer security and software engineering venues.

A cornerstone of fuzzing research, and science in general, is that other researchers can critically assess the correctness of scientific results. 
To this end, the research results must be \emph{reproducible}, meaning that another group should be able to obtain the same results using the same experimental setup, often by using a research artifact provided by the authors~\cite{ACM.artifact.20}.
Reproducibility is paramount for other researchers to understand, trust, and build on the research results.

To enable high-quality research and provide a common foundation for evaluating fuzzing methods, several works describe how newly proposed fuzzing approaches should be evaluated.
In 2018, the first and most influential paper describing a reproducible evaluation design was published by Klees~\etal~\cite{klees2018evaluating}. %
It describes guidelines to advise researchers on how fuzzing research should evaluate their respective contributions. 
For example, a crucial insight introduced by Klees~\etal is the repetition of experiments to account for the inherent randomness of the fuzzing process. 
Although Klees~\etal recommend ``a sufficient number of trials'' and use 30 trials in their own experiments, we found that in practice, this recommendation is interpreted as anything between three and 20 repetitions. Another guideline is to confirm the fuzzers' performance statistically; however, this makes little sense with few repetitions and is often skipped.

In this work, we systematically review how the recommendations for evaluating fuzzing methods are implemented in practice and critically evaluate the reproducibility of fuzzing research. 
We propose revised best practices for evaluating fuzzing methods and point out pitfalls that we have observed in practice.
In other fields, such work has had a significant impact on improving research from a methodological point of view~\cite{abadi1996prudent,vanderkouwe2019benchmarkingcrimes, demir2022reproducibility, arp2022dosanddonts}.

We conduct a thorough literature review of \numpapers fuzzing papers published in prestigious A$^*$ venues---as ranked by CORE2023~\cite{coreranking}---between 2018 and 2023. 
While we primarily focus on computer security venues, namely IEEE Symposium on Security and Privacy (S\&P), USENIX Security Symposium (USENIX), ACM Conference on Computer and Communications Security (CCS), and ISOC Network and Distributed System Security (NDSS) Symposium, we also examine three software engineering venues: IEEE/ACM International Conference on Automated Software Engineering (ASE), ACM Joint European Software Engineering Conference and Symposium on the Foundations of Software Engineering (ESEC/FSE), and International Conference on Software Engineering (ICSE). 
For all papers, we:
\begin{inparaenum}[(i)]
\item systematically analyze how evaluations are conducted (in terms of metrics, targets, baselines, reported bugs, etc.),
\item check whether common fuzzing guidelines (as outlined by Klees~\etal~\cite{klees2018evaluating} or embodied in implicit community wisdom, \eg ``do not use artificial bug datasets'') are followed, and
\item investigate potential flaws threatening the validity of the respective evaluation.
\end{inparaenum}

Following our literature analysis, we present \numcasestudies case studies of fuzzing papers across different fields and attempt to reproduce (parts of) their evaluation. 
For each case study, we discuss any shortcomings we have identified because they illustrate potential pitfalls of which researchers should be aware.
Note that these case studies are \emph{not} intended to point fingers or criticize any particular work. 
Instead, we aim to highlight potential challenges that can affect the outcome of a research paper and explore what aspects need to be considered when designing the evaluation of a fuzzing method.
Based on the findings of our literature review and case studies, we propose best practices for evaluating future fuzzing methods to enable reproducible research. 

\smallskip \noindent
In summary, we make the following key contributions:

\begin{itemize}
\item We conduct a systematic literature survey of \numpapers papers published in the past six years at top venues to assess how fuzzing methods are typically evaluated.
\item We attempt to reproduce \numcasestudies papers to assess the practical aspect of fuzzing evaluations. In doing so, we identify several obstacles that illustrate (sometimes subtle) shortcomings of evaluating fuzzing methods.
\item Based on our lessons learned, we provide revised recommendations and best practices for future fuzzing evaluations.
\end{itemize}

Supplementary material for this work is available online at \url{https://github.com/fuzz-evaluator/}, including our reproduction artifacts and recommended best practices for future work (see \url{https://github.com/fuzz-evaluator/guidelines}).

\section{Fuzzing Evaluation Guidelines}\label{sec:guidelines}

We first provide a brief overview of fuzzing before describing several generally accepted best practices that guide a typical fuzzing evaluation.

\subsection{Background on Fuzzing}

Fuzzing, also referred to as \emph{fuzz testing}, is a dynamic testing technique with the goal of uncovering bugs in systems. 
This typically happens by mutating some initial input(s) to the system or by deriving inputs from input specifications such as grammars. While processing the provided input, the system under test is monitored for \emph{interesting} behavior. Beyond easily observable faults, such as program crashes, fuzzers can use more sophisticated bug oracles, such as sanitizers or differential testing. 
Moreover, modern fuzzers often use lightweight instrumentation to receive coverage feedback, allowing them to track inputs that executed previously unseen edges. 
A comprehensive overview of various fuzzing techniques can be found in the \emph{Fuzzing Book}~\cite{zeller2019fuzzingbook}, and several surveys present a comprehensive overview of this topic~\cite{manes2021survey,zhu2022survey} or open challenges in this domain~\cite{boehme2021challenges}.
Most fuzzing research proposes an improvement by way of new techniques, new components, or entirely new fuzzers---few works focus on the theory behind fuzzing~\cite{boehme2020_exponentialcost, boehme2021_residualrisk, boehme2022_covreliability, liyanage2023reachablecoverage}.

A fundamental principle of all fuzzers is the inherent inclusion of randomness into the testing process. Starting from the scheduling order of the process, through the input and the mutations applied to it, to the fuzzing environment (including functions such as \texttt{getpid}, \texttt{time}, or \texttt{rand}, or shared resources such as the filesystem), there are numerous sources of randomness that make deterministic and reproducible execution challenging.

\subsection{Guidelines of \emph{Evaluating Fuzz Testing}}

The randomized nature of fuzzing needs to be taken into account during the evaluation, which leads to challenges with reproducibility of research results in practice. Hence, the seminal paper by Klees~\etal~\cite{klees2018evaluating} outlined several guidelines on how a proper fuzzing evaluation should be conducted. For a reproducible and fair evaluation, they propose the following recommendations:

\noindent\textbf{\bfseries Recommendation\hspace{0.3em}1 -- Baseline:} A comparison with a relevant and reasonable baseline is imperative to show what improvement a particular fuzzer provides.

\noindent\textbf{\bfseries Recommendation\hspace{0.3em}2 -- Targets:} A relevant sample of targets to compare against is necessary. This includes benchmark programs with known bugs that can be used as a ground truth to measure bug detection capabilities.

\noindent\textbf{\bfseries Recommendation\hspace{0.3em}3 -- Setup \& Parameters:} Due to the inherent randomness of fuzzing, individual runs with the same configuration can yield significantly different outcomes. To address this problem, Klees~\etal propose repeating the experiment multiple times. Similarly, fuzzing performance may vary within a single run, so short runtimes are not appropriate for extrapolating the behavior of a fuzzer over longer times. They propose 24 hours as a reasonable fuzzer runtime and recommend plotting the performance over time. Seed sets must be well documented and carefully selected; ideally, various sets, including the empty or uninformed seed, are tested.

\noindent\textbf{\bfseries Recommendation\hspace{0.3em}4 -- Evaluation Metrics:} Ideally, fuzzing evaluations should not be based on proxy metrics such as code coverage alone, but on a fuzzer's ability to find bugs, \ie the goal for which it was designed. In particular, an evaluation should not rely on heuristics such as AFL's coverage profiles or stack hashing. Complementing the evaluation on bug detection, Klees~\etal recommend code coverage in terms of basic blocks or edges as secondary metric.

\noindent\textbf{\bfseries Recommendation\hspace{0.3em}5 -- Statistical Evaluation:} Finally, the fuzzing evaluation should undergo statistical evaluation to rule out that the observed behavior is by mere chance. This requires a \emph{sufficient} number of trials (Klees~\etal themselves use 30); then, a statistical test such as the Mann-Whitney U-test or bootstrap-based methods should be used to test the null hypothesis that the new method exhibits no difference compared to a reasonable baseline.

\subsection{Guidelines of \emph{\fuzzbench}}

\fuzzbench~\cite{metzman2021fuzzbench}, a benchmarking suite for general-purpose fuzzer evaluation developed by Google, provides several target programs and aims to provide a standardized setup for fair comparison of fuzzers. \fuzzbench is the successor to the Google Fuzzer Test Suite (FTS)~\cite{google2016fts}.
During their extensive evaluation, the authors made two key observations that can serve as a guideline for future fuzzing research. 
First, the performance of a fuzzer varies significantly depending on the number of initial seeds; running without seeds allows for studying the difference when only a particular fuzzer can solve some comparisons/branches.
Second, using a saturated corpus for fuzzing is \emph{not} recommended, as fuzzers are barely capable of augmenting it. Even though this is common in practice, it is not well suited to discern or measure the performance of fuzzers.

\subsection{Guidelines of \emph{On the Reliability of Coverage}}

More recently, Böhme~\etal~\cite{boehme2022_covreliability} made a number of recommendations based on their evaluation of the reliability of coverage. In particular, they recommend to use at least ten representative programs, each tested at least ten times for at least 12 hours (preferably, each value should be doubled). The selected programs should be real-world programs, and a bug evaluation should be done on real-world bugs. In addition to bugs, code coverage should also be evaluated---both using established metrics. In particular, fuzzer-specific measures such as AFL's unique paths should be avoided. For comparison, authors should choose a suitable baseline, such as the fuzzer on top of which the new technique is implemented. Authors should consider splitting benchmarks into a \emph{training} and \emph{validation} set to avoid overfitting. To confirm evaluation results, authors must measure significance and effect size using established techniques. They should discuss threats to the validity of their evaluation and how they mitigated them. Finally, authors should carefully document their setup and publish evaluation artifacts on long-term stable platforms such as Zenodo.

\subsection{Fuzzing Benchmarks}

Over the years, several \emph{standardized benchmarks} and \emph{platforms} to conduct fair and comparable fuzzing evaluations have been proposed, \eg 
\fts~\cite{google2016fts} (2016; superseded by \fuzzbench),
\lavam~\cite{dolan2016lavam} (2016),
\cgc~\cite{darpa2018cgc} (2018),
\magma~\cite{hazimeh2020magma} (2020),
\fuzzbench~\cite{metzman2021fuzzbench} (2020), 
\unibench~\cite{li2021unifuzz} (2021), 
\profuzzbench~\cite{natella2021profuzzbench} (2021), and %
\revbugbench~\cite{zhang2022fixreverter} (2022).

These benchmark platforms aim to measure the performance of general-purpose fuzzing, except for \profuzzbench, which focuses on stateful protocol fuzzing. Overall, we can distinguish between benchmarks focusing on the comparison of achieved coverage (\fts, \unibench, \fuzzbench, and \profuzzbench) and those focusing on the bug-finding capabilities of the fuzzing technique (\lavam, \cgc, \magma, and \revbugbench). In the latter category, some utilize artificial bug injection (\lavam and \cgc), make efforts to port actual vulnerabilities to the latest version of a program (\magma), or to revert fixes (\revbugbench).
Artificial bug injection methods often introduce shallow bugs that are amenable to fuzzers, and are generally no longer recommended for an evaluation~\cite{metzman2021fuzzbench,bundt2021evaluating,wang2020tortoisefuzz,zhang2022fixreverter}.

\begin{table}[tb]
    \caption{Overview of analyzed papers.}%
    \label{tab:paper_overview}
    \centering
    \resizebox{\columnwidth}{!}{%
    \begin{tabular}{lllr}
    \toprule
    \multicolumn{1}{c}{\textbf{Year}} & \multicolumn{1}{c}{\textbf{Venue}} & \multicolumn{1}{c}{\textbf{Papers}} & \multicolumn{1}{c}{\textbf{Studied}} \\
    \midrule

    \multirow{7}{*}{2023} & ASE$^*$ & \cite{wang2023mlirsmith}, \cite{humayun2023naturalfuzz}, \cite{liu2023vd-guard} & 3/7 \\
                          & FSE$^*$ & \cite{wu2023sjfuzz} & 1/6 \\
                          & ICSE & \cite{jiang2023cofuzz}, \cite{jia2023jopfuzzer}, \cite{wu2023jitfuzz}, \cite{guo2023typeoracle}, \cite{lee2023seamfuzz} & 5/11 \\
                          & CCS & \cite{zhang2023_profopt}, \cite{deng2023nestfuzz}, \cite{chen2023hopper}, \cite{meng2023mallory} & 4/9 \\
                          & NDSS & \cite{jauernig2023darwin}, \cite{groß2023fuzzilli}, \cite{bulekov2023fuzzng} & 3/4 \\
                          & S\&P & \cite{luo2023selectfuzz}, \cite{busch2023teezz}, \cite{liu2023videzzo} & 3/9 \\
                          & \multirow{2}{*}{USENIX} & \cite{shi2023aifore}, \cite{luo2023bleem}, \cite{zheng2023fishfuzz}, \cite{seidel2023safirefuzz}, \cite{christou2023ivysyn}, \cite{lyu2023miner}, \\
                          &  & \cite{stone2023_nolinuxnoproblem}, \cite{angelakopoulos2023firmsolo}, \cite{bars2023fuzztruction}, \cite{wang2023fuzzjit}, \cite{peng2023gleefuzz}, \cite{li2023polyfuzz} & \multirow{-2}{*}{12/29} \\
    \midrule
    \multirow{7}{*}{2022} & ASE & \cite{fu2022griffin}, \cite{yu2022htfuzz} & 2/4 \\
                          & FSE & \cite{gu2022glibfuzzer}, \cite{zhou2022minerva} & 2/6 \\
                          & ICSE & \cite{li2022microafl}, \cite{nguyen2022bedivfuzz}, \cite{kukucka2022confetti}, \cite{wei2022freefuzz}, \cite{green2022graphfuzz}, \cite{meng2022ltl-fuzzer}, \cite{song2022r2z2}, \cite{du2022windranger} & 8/17 \\
                          & CCS & \cite{jiang2022evocatio}, \cite{bernhard2022jit-picker}, \cite{fioraldi2022libafl}, \cite{shah2022mc2}, \cite{chen2022metaemu}, \cite{chen2022sfuzz}, \cite{zhou2022semu} & 7/8 \\
                          & NDSS & \cite{jiang2022conzzer}, \cite{xu2022cooper}, \cite{zhang2022mobfuzz} & 3/6 \\
                          & S\&P & \cite{huang2022beacon}, \cite{she2022k-scheduler}, \cite{lin2022grebe}, \cite{chen2022jigsaw}, \cite{liang2022pata} & 5/9 \\
                          & \multirow{2}{*}{USENIX} & \cite{shen2022drifuzz}, \cite{zhou2022ferry}, \cite{zhang2022fixreverter}, \cite{scharnowski2022fuzzware}, \cite{myung2022mundofuzz}, & \\
                          & & \cite{cloosters2022sgxfuzz}, \cite{zhao2022statefuzz}, \cite{ba2022sgfuzz}, \cite{chen2022symsan} & \multirow{-2}{*}{9/19} \\
    \midrule
    \multirow{7}{*}{2021} & ASE & \cite{liu2021instruguard}, \cite{jiang2021rulf} & 2/6 \\
                          & FSE & \cite{metzman2021fuzzbench}, \cite{zhang2021heterofuzz} & 2/4 \\
                          & ICSE & \cite{borzacchiello2021fuzzy-sat}, \cite{vikram2021_bonsaifuzzing}, \cite{park2021jest} & 3/6 \\
                          & CCS & \cite{ge2021hyperfuzzer}, \cite{zhu2021aflchurn}, \cite{nagy2021hexcite}, \cite{feng2021snipuzz}, \cite{he2021sofi}, \cite{chen2021syzgen} & 6/13 \\
                          & NDSS & \cite{dinh2021favocado}, \cite{poeplau2021symqemu}, \cite{jung2021winnie-afl} & 3/6 \\
                          & S\&P & \cite{mera2021dice}, \cite{choi2021ntfuzz} & 2/7 \\
                          & USENIX & \cite{lee2021cafl}, \cite{schumilo2021nyx}, \cite{fioraldi2021invscov}, \cite{salls2021token-level-afl} & 4/13 \\
    \midrule
    \multirow{7}{*}{2020} & ASE & \cite{nguyen2020mofuzz}, \cite{zhou2020zeror} & 2/4 \\
                          & FSE & \cite{boehme2020entropic}, \cite{song2020crfuzz}, \cite{she2020mtfuzz} & 3/7 \\
                          & ICSE & \cite{manès2020ankou}, \cite{wen2020memlock}, \cite{wuestholz2020bran}, \cite{wang2020uafl} & 4/6 \\
                          & CCS & \cite{xu2020freedom} & 1/2 \\
                          & NDSS & \cite{kim2020hfl}, \cite{schumilo2020hyper-cube}, \cite{wang2020tortoisefuzz} & 3/4 \\
                          & S\&P & \cite{park2020die}, \cite{aschermann2020afl-ijon}, \cite{xu2020krace}, \cite{huang2020pangolin}, \cite{dinesh2020retrowrite}, \cite{chen2020savior} & 6/7 \\
                          & \multirow{2}{*}{USENIX} & \cite{song2020agamotto}, \cite{yue2020ecofuzz}, \cite{ruge2020frankenstein}, \cite{ispoglou2020fuzzgen}, \cite{gan2020greyone}, \cite{clements2020halucinator},  \\
                          & & \cite{chen2020muzz}, \cite{lee2020montage}, \cite{feng2020p2im}, \cite{oesterlund2020parmesan}, \cite{poeplau2020symcc} & \multirow{-2}{*}{11/19} \\
    \midrule
    \multirow{7}{*}{2019} & ASE & \textcolor{black!40}{--} & \textcolor{black!40}{0/0} \\
                          & FSE & \cite{li2019cerebro} & 1/4 \\
                          & ICSE & \cite{chen2019classming}, \cite{nilizadeh2019diffuzz}, \cite{choi2019eclipser}, \cite{you2019slf}, \cite{wang2019superion} & 5/7 \\
                          & CCS & \cite{cho2019intriguer}, \cite{chen2019matryoshka} & 2/3 \\
                          & NDSS & \cite{han2019codealchemist}, \cite{aschermann2019nautilus}, \cite{aschermann2019redqueen} & 3/4 \\
                          & S\&P & \cite{nagy2019untracer}, \cite{xu2019janus}, \cite{she2019neuzz}, \cite{you2019profuzzer}, \cite{jeong2019razzer} & 5/5 \\
                          & USENIX & \cite{gueler2019antifuzz}, \cite{chen2019enfuzz}, \cite{zheng2019firm-afl}, \cite{jung2019fuzzification}, \cite{blazytko2019grimoire}, \cite{lyu2019mopt} & 6/6 \\
    \midrule
    \multirow{7}{*}{2018} & ASE & \cite{lemieux2018fairfuzz} & 1/2 \\
                          & FSE & \textcolor{black!40}{--} & \textcolor{black!40}{0/0} \\
                          & ICSE & \textcolor{black!40}{--} & \textcolor{black!40}{0/0} \\
                          & CCS & \cite{chen2018hawkeye} & 1/2 \\
                          & NDSS & \cite{chen2018iotfuzzer}, \cite{muench2018wycinwyc} & 2/2 \\
                          & S\&P & \cite{gan2018collafl}, \cite{peng2018t-fuzz} & 2/3 \\
                          & USENIX & \cite{talebi2018charm}, \cite{pailoor2018moonshine}, \cite{yun2018qsym} & 3/3 \\
    \midrule
    \multicolumn{3}{l}{\textit{total \#papers analyzed}} & 150/289 \\
    \bottomrule
    \\[-0.8em]
    \multicolumn{4}{l}{$^*$ \emph{limited to available preprints}} 
    \end{tabular}
    }
\vspace{-1.4em}
\end{table}

\section{Literature Analysis}\label{sec:literature_analysis}

With these guidelines and benchmarks in mind, we now study their adoption to better understand what best practices are used in fuzzing research.
To this end, we perform a comprehensive literature survey of recent fuzzing papers. %

\subsection{Method}\label{sec:method}

We examine all fuzzing papers published at the top computer security and software engineering conferences between 2018 and 2023\footnote{For 2023, ASE and FSE have not published the papers at the time of writing. We therefore work with available preprints.}. %
We include a paper in our analysis if its focus is on fuzzing, \eg it proposes a new method or extensively evaluates existing ones. In contrast, we exclude papers using fuzzers as a means to support their primary focus, \eg solely to generate some diverse inputs.
We identify \totalpapers candidate papers for which we collect metadata about the underlying evaluation method, including whether the paper successfully participated in an artifact evaluation process. We then randomly select \numpaperswperc from these \totalpapers papers and manually review them, \ie study the design and evaluation of the work in detail. Table~\ref{tab:paper_overview} shows an overview of analyzed papers.

We investigate whether the fuzzing evaluation guidelines outlined in Section~\ref{sec:guidelines} are followed or whether an evaluation deviates from them. We want to stress that there may be good reasons to deviate from these guidelines, making a manual review and judgment on a case-by-case basis mandatory. We also study whether the evaluations performed expose flaws that future fuzzing papers could avoid.

\subsection{Results}\label{sec:results}
We study the papers regarding their reproducibility, targets, fuzzers, evaluation setup in terms of resources, common metrics, and statistical evaluation.

\subsubsection{Reproducibility}

A crucial aspect of verifying and advancing science is the ability to reproduce existing research results. When examining the metadata we collected for all \totalpapers fuzzing papers, we find that \numcodeavailablewperc publish the code of their technique, while \numcodeunavailablewperc do not share their code. Some do not contribute new code, upstreamed their code, or have not yet released the code (applies to FSE, which will take place after time of writing). 
Regarding other data (excluding code), we find that \numdataavailablewperc share data, 
\numdataavailableinsteadofcode of which publish data as a substitute because they do not share their code or have no code to share. 
All software engineering conferences (ASE, FSE, and ICSE),  USENIX Security, and CCS (since 2023) offer an artifact evaluation process where independent reviewers assess the published research artifact (for 2023, ASE and FSE have not yet published this data). Our analysis found that \numaeunavailablewperc of the papers did not have access to such an artifact evaluation; \numaenobadgewperc had access but opted to not participate or failed to receive any badge. Only \numaeanybadgewperc of the papers have one or more badges. Of these, \numaeavailable are considered \emph{available} and \numaefunctional \emph{functional} or \emph{reusable}, a crucial requirement for reproduction. USENIX Security and CCS offer to reproduce the results of a paper, which only \numaereproduced out of \numaeaccesstoreproduced eligible papers achieved. We emphasize that artifact evaluation has been introduced only in recent years, but participation is rising. 
CCS offered artifact evaluation for the first time in 2023, further supporting this trend.
\begin{mdframed}[style=insightstyle]
With \numcodeavailableperconly, a majority of works releases their code. Despite being relatively new, \numaeaccessperconly of the papers already had access to artifact evaluation, with adoption lagging behind at \numaeanybadgeperconly of papers that obtained a badge.
\end{mdframed}

\subsubsection{Targets under Test}
\begin{figure}[tb]
    \centering
    \includegraphics[width=\columnwidth]{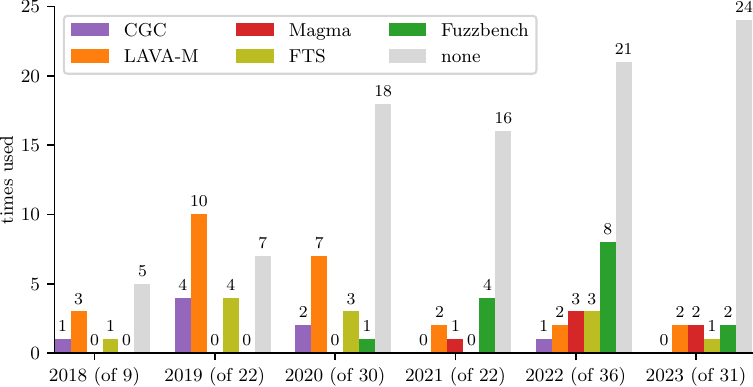}
    \vspace{-1.5em}
    \caption{Benchmark usage over the years. The numbers in brackets represent the number of papers analyzed for the respective year. Note that some papers use multiple benchmarks, hence the numbers do not add up.}%
    \label{fig:dataset_usage}
\end{figure}

To showcase the strengths of an approach, a suitable set of targets is required. Looking at the distribution of used targets (excluding datasets) in Table~\ref{tab:targets}, we find that they are strongly biased towards byte-oriented file formats, especially binutils. On average, fuzzing papers evaluate on \numavgtargets targets. In summary, we found \numuniquetargets different targets used across all studied papers; of these, \numtargetsmaxonceusedwperc were evaluated in only one paper. %
In addition to real-world targets, a common way of reproducibly measuring fuzzer performance is using benchmarks. Figure~\ref{fig:dataset_usage} shows how benchmarks have been adopted in the past years. In total, \numdatasetusednonewperc of the papers use no benchmark, \numdatasetusedlavamwperc use LAVA-M~\cite{dolan2016lavam}, \numdatasetusedfuzzbenchwperc use \fuzzbench~\cite{metzman2021fuzzbench}, \numdatasetusedftswperc use Google's Fuzzer Test Suite (FTS)~\cite{google2016fts}, \numdatasetusedcgcwperc DARPA's CGC binaries (CGC)~\cite{darpa2018cgc}, \numdatasetusedmagmawperc rely on \magma~\cite{hazimeh2020magma}, and \numdatasetusedunifuzzwperc build on \unibench~\cite{li2021unifuzz} for benchmarking purposes.
Despite its success, \lavam is nowadays considered flawed because it artificially injects vulnerabilities into a given target program that are easy for a fuzzer to find but do not correspond to real bugs~\cite{metzman2021fuzzbench,bundt2021evaluating,wang2020tortoisefuzz,zhang2022fixreverter}. More recent works using \lavam often do so only for comparability reasons~\cite{jauernig2023darwin,jiang2023cofuzz}. Similar to \lavam, \cgc is widely considered outdated and inadequate.

\begin{mdframed}[style=insightstyle]
Real-world targets are often limited to binary input-affine programs, while benchmarks are not used by the majority of papers. Benchmarks with artificial vulnerabilities are still used.
\end{mdframed}

\begin{table}[tb]
    \centering
    \caption{Targets fuzzed in five or more analyzed papers (excluding benchmarks). Some papers report generically to evaluate on binutils, while others specify exact targets, such that numbers in practice may differ slightly.}%
    \label{tab:targets}
    \resizebox{\columnwidth}{!}{%
    \begin{tabular}{rl}
    \toprule
    \textbf{\#Uses} & \textbf{Target} \\
    \midrule
    25 & objdump, readelf \\
    20 & nm, tcpdump \\
    19 & libpng \\
    17 & libtiff \\
    13 & cxxfilt, jhead, libjpeg \\
    12 & libxml2 \\
    11 & nasm \\
    10 & jasper, libming, openssl, size \\
     9 & file, ImageMagick, mjs, tiff2pdf \\
     8 & djpeg, exiv2, JavaScriptCore, libarchive, SQLite, v8, xmllint \\ 
     7 & ChakraCore, ffmpeg, harfbuzz \\
     6 & binutils, lcms, lrzip, mupdf, OpenJPEG, SpiderMonkey  \\
     \multirow{2}{*}{5} & bento, bsdtar, catdoc, cflow, curl, freetype2, GraphicMagick,\\
       & json, pcre2, proj4, strip, tiff2ps, yara, zlib \\
     \bottomrule
    \end{tabular}
    }%
\end{table}

\subsubsection{Evaluation against State of the Art}
Comparison with a strong set of existing work helps to demonstrate that a new method is particularly suited to solve a specific problem. Yet, only a few techniques published in the past few years have been broadly incorporated in follow-up work. 
Instead, the most famous fuzzers extended with new techniques are \afl~\cite{afl} with \numbasedonaflwperc, AFL++~\cite{aflplusplus} with \numbasedonaflppwperc, libFuzzer~\cite{libfuzzer} with \numbasedonlibfuzzerwperc, and syzkaller~\cite{syzkaller} with \numbasedonsyzkallerwperc. 
Interestingly, all of these tools are non-academic works; only for AFL++ a peer-reviewed paper has been published~\cite{aflplusplus}. Contrasting this number, \numbasedonnonewperc of the proposed tools are not based on any existing tool.

When looking at the fuzzers chosen as baselines for comparison, we find that \afl is compared against by \numcompetitoraflwperc of studies, followed by QSym~\cite{yun2018qsym} with \numcompetitorqsymwperc, AFLFast~\cite{boehme2017aflfast} with \numcompetitoraflfastwperc, Angora~\cite{chen2018angora} with \numcompetitorangorawperc, FairFuzz~\cite{lemieux2018fairfuzz} with \numcompetitorfairfuzzwperc, and AFL++ with \numcompetitoraflppwperc.
From the \numpapers papers we analyzed, only QSym (2018), FairFuzz (2018), and MOpt~\cite{lyu2019mopt} (2019) have been chosen by more than five follow-up works for comparison. 
More recently, only Fuzzilli~\cite{groß2023fuzzilli} (published 2023, open-sourced early 2019) 
was used by multiple works for their evaluation, even before the paper was published.
This does not account for techniques replicated in AFL++ or LibAFL~\cite{fioraldi2022libafl}, which reimplement many successful techniques proposed~\cite{boehme2017aflfast, aschermann2019redqueen, lafintel, lyu2019mopt}. On average, a fuzzing paper evaluates against \numavgcompetitors other fuzzers.

Analyzing whether papers omit comparing against a relevant fuzzer in their evaluation, we find that \nummissingcompetitorsmissingtoolswperc of the works ignore at least one relevant state-of-the-art method and \nummissingcompetitorsdoesnotcompareagainstbaselinewperc even omit comparing against their baseline, \ie the tool on which they base their own fuzzer.

\begin{mdframed}[style=insightstyle]
45\% of fuzzing research builds on top of non-academic fuzzers, \numbasedonnoneperconly build a new tool. 23\% percent of fuzzing evaluations fail to compare against relevant state-of-the-art fuzzers or their own baseline.
\end{mdframed}

\subsubsection{Evaluation Setup}
With respect to the evaluation setup, we analyze the runtime, the number of CPU cores assigned, whether all resources were allocated fairly, and the seeds used for the experiments.

\begin{figure}[b]
   \centering
   \graphicspath{{graphics}}
   \def\svgwidth{\columnwidth}
   \begin{scriptsize}
       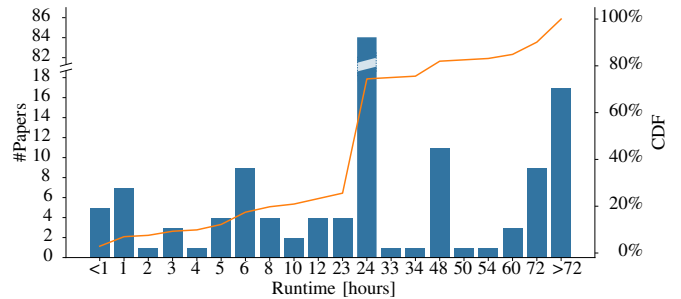
   \end{scriptsize}
   \vspace{-1.5em}
   \caption{Distribution of runtimes used in practice and cumulative distribution function (CDF), which shows that \numruntimelessthantwentythreeperconly of papers use a runtime of less than 23 hours. \numpapersmultipleruntimes papers use multiple, different runtimes; we include all in these cases.}%
   \label{fig:runtime_distribution}
\end{figure}

\paragraph{\bfseries Runtime} Reviewing the experiment setup used across fuzzing evaluations, we find that the majority of papers uses a runtime of 24h, more precisely \numruntimetwentyfourhourswperc of the papers run at least one experiment for 24 hours. As Figure~\ref{fig:runtime_distribution} outlines, only \numruntimelessthantwentythreewperc of the works use a runtime of less than 23 hours, while \numruntimemorethantwentyfourwperc use an even higher runtime. \numruntimenonewperc do not specify their runtime or have no own experiments measuring time.

\paragraph{\bfseries CPU cores} In terms of CPU cores assigned to fuzzers, we find an inconsistent picture, with a significantly varying number of CPU cores used. The most common result was that \numcoresnonewperc of the papers did not specify how many CPU cores they used, \numcoresonewperc used one core, and \numcorestwowperc used two cores.

\paragraph{\bfseries Fair computing resources} When checking whether the available computing resources were allocated fairly (\eg the same number of cores were allocated to each fuzzer and they were run for the same amount of time), we find that this is the case for \numresourceallocationfairwperc of the works. For \numresourceallocationnotspecifiedwperc, we could not infer this information from the description in the paper, and \numresourceallocationnotapplicablewperc did not evaluate other fuzzers or did not conduct any experiments where this was an issue. Crucially, \numresourceallocationunfairwperc unfairly allocate resources, giving one fuzzer an advantage over another.
For these \numresourceallocationunfair, we found one benign case in which an existing method was given more resources, one case in which the number of executions was fairly distributed rather than the runtime (thereby giving slow fuzzers an advantage), two cases in which a different number of cores was used (in one case, giving the new fuzzer twice the cores than others), and four cases where the new approach was allowed some preprocessing time, \eg for some static analysis pass or seed preprocessing, before it was then allotted the same time as all other tools, effectively giving it more computation time. Unfortunately, the authors rarely explain their motivation for doing so, nor do they consider consequences for the evaluation.
Also, our analysis does not address manual work, which may be distributed unfairly between different fuzzers, for example, giving one fuzzer a fine-tuned configuration that performs better.

\paragraph{\bfseries Initial seeds} Another crucial factor determining a fuzzer's performance is the set of initial seeds~\cite{klees2018evaluating,herrera2021seed}. We studied if the \emph{type} of seeds is specified and if information on concrete seed files is available. 
Out of the \numpapers papers, \numuninformedseedsnoseedsrequiredwperc require no seeds, \numuninformedseedsuninformedseedswperc use uninformed or empty seeds, \numuninformedseedsinformedseedswperc use informed seeds, \numuninformedseedsbenchmarkseedswperc use seeds provided by the project as test cases or those that are shipped with a benchmark, and \numuninformedseedsmultipleseedsetswperc use multiple types of seed sets, while \numuninformedseedsnotspecifiedwperc do not specify at all what type of seeds are used, making a reproduction challenging. 
Regarding concrete details, we find that \numseedexplanationnoinfoavailablewperc of the papers fail to disclose what seeds they use, compared to \numseedexplanationinfoonseedsavailablewperc that outline their seeds. %
A further pitfall potentially threatening an evaluation's validity is the fair distribution of the same seeds to all fuzzers. While this is the case in \numseedfairnesssameseedsforallwperc of the studied papers, in \numseedfairnessnotspecifiedwperc of the works this does not become clear, and \numseedfairnessdifferentseedswperc even use diverging seed sets. Three of these cases arise due to the fuzzer design or other fuzzers lacking the capability to process a particular type of input. We stress that this may be valid, for example, when a fuzzer used for comparison needs a larger seed set than the proposed fuzzer, yet giving a fuzzer a different set of seeds requires special attention and documentation. 

\begin{mdframed}[style=insightstyle]
We find that \numresourceallocationunfairperconly of the papers allocate computing resources unfairly, and \numseedfairnessdifferentseedsperconly use different seed sets.
\end{mdframed}

\subsubsection{Evaluation Metrics}
While many different metrics exist, often specific to the particular technique introduced, a small number of metrics has found widespread adoption: \numevalmetricscodecoveragewperc of the papers use some sort of \emph{code coverage}, and \numevalmetricsbugswperc use the (re-)discovery of bugs as a metric to compare fuzzers. The third most widespread metric, Time-To-Exposure (TTE), is used by \numevalmetricsttewperc of the papers, mainly from the directed fuzzing domain.

\paragraph{\bfseries Code Coverage} Code coverage comes in different forms; the most popular are the following: \numcoveragebranchcoveragewperc of the papers use branch coverage, \numcoverageedgeswperc employ edge coverage, \numcoveragebasicblockswperc rely on basic block coverage, and \numcoveragelinecoveragewperc use line coverage on the source code level. Furthermore, \numcoveragepathswperc use some notion of paths to measure coverage. We stress this metric is unreliable \emph{without} a definition of what the paper considers a path. Differences exist, for example, between actual program paths and AFL's path metric, requiring any paper to specify what they consider a path for their work. 
Beyond the type of coverage, the process of measuring coverage is also prone to errors, and the concrete choice of measurement is often not documented. In total, we find that \numcoverageexplanationnodefinitionexplanationwperc of the works lack a clear definition or explanation of how they measure coverage, whereas \numcoverageexplanationyeswperc document this (the remaining papers do not measure coverage). 
For example, measuring coverage using a binary with instrumentation that not all fuzzers had access to during the fuzzing campaign gives some fuzzers an advantage.
Similarly, when measuring coverage on a bitmap with collisions, the reported coverage is up to 9\% smaller~\cite{lipp2022fuzztastic} than the true one. This may cause problems when a different bitmap size was used during fuzzing, as the inputs saved by a fuzzer may no longer trigger the new coverage on the bitmap with collisions.
A further pitfall affects emulation-based fuzzing, especially when using QEMU~\cite{qemu2005bellard}. We observed that papers often provide no clear distinction between translated blocks as presented by the emulator and actual basic blocks for the target binary. We found that in at least one case this led to overcounting the reached coverage, as translated blocks were mistaken for basic blocks.

\paragraph{\bfseries Known Bugs} As research from Klees~\etal~\cite{klees2018evaluating} as well as Böhme~\etal~\cite{boehme2022_covreliability} points out, coverage may not be an accurate proxy for bug finding, even though a strong correlation exists. Ultimately, a fuzzer's goal is finding bugs, making the evaluation of whether it can find known or unknown vulnerabilities an excellent experiment. Known bugs are a good way of measuring a fuzzer's performance, yet it is difficult to find suitable bugs outside well-designed benchmarks, such as Magma~\cite{hazimeh2020magma} or RevBugBench~\cite{zhang2022fixreverter}. 

\paragraph{\bfseries New Bugs / CVEs} Another commonly used approach is the capability of finding previously unknown bugs. 
Ethical handling requires researchers to responsibly disclose these bugs to the vendors or maintainers. Both sides can additionally request a CVE that serves as a unique identifier for the found vulnerability.
In practice, CVEs have become a commonly used metric to assess whether a fuzzer can find bugs in real-world software, presumably showing its impact. Of the \numpapers analyzed papers, \numallpaperswithcves claim one or more CVEs (\numavgcvesclaimed on average, \numallcves in total). Given the implicit expectation of submissions to have a real-world impact, the authors often try to obtain as many CVEs as possible. We randomly selected \numpaperscveanalysis of these papers~\cite{he2021sofi, ispoglou2020fuzzgen, lyu2019mopt, wang2019superion, liu2021instruguard, busch2023teezz, jiang2023cofuzz, li2022microafl, zhang2021stochfuzz, chen2019enfuzz, choi2021ntfuzz, dinh2021favocado, chen2019classming, zheng2023fishfuzz, zhou2022minerva, yu2022htfuzz, pan2021v-shuttle, li2023polyfuzz, deng2023nestfuzz, chen2021syzgen, she2019neuzz, liu2023vd-guard, christou2023ivysyn, luo2023bleem, yue2020ecofuzz, myung2022mundofuzz, wen2020memlock, meng2022ltl-fuzzer, salls2021token-level-afl, zhao2022statefuzz, ba2022sgfuzz, huang2020pangolin, lee2020montage, pailoor2018moonshine, du2022windranger} and analyze the \numcvesanalyzed CVEs they claim (\numcvesanalyzedperconly of all CVEs claimed across the \numallpaperswithcves papers).

\begin{figure}
    \centering
    \graphicspath{{graphics}}
    \def\svgwidth{\columnwidth}
    \begin{scriptsize}
         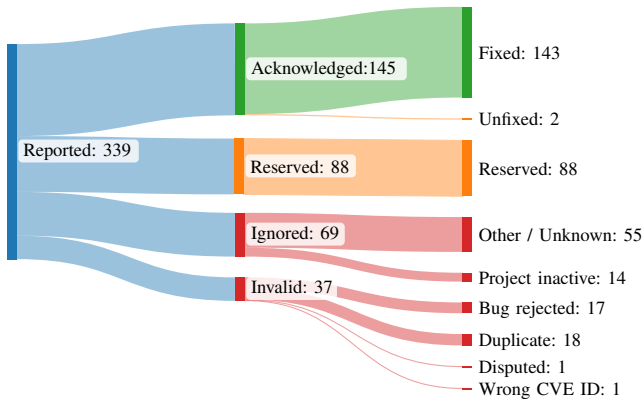
    \end{scriptsize}
    \caption{Outcome of \numcvesanalyzed CVEs that were reported across \numpaperscveanalysis papers. Only \numcvecategoryfixedperconly of the CVEs have been acknowledged by the developers. Pending public disclosure, information on CVEs in the \emph{Reserved} state is withhold.}%
    \label{fig:cve_breakdown}
\end{figure}

As Figure~\ref{fig:cve_breakdown} shows, surprisingly, only \numcvecategoryfixedwperc of the CVEs are valid (\ie neither formally disputed, reserved, nor ignored or rejected by the project maintainers) and have been fixed (or at least acknowledged). \numcvecategoryblindedwperc of the CVEs were still marked as \texttt{RESERVED}, preventing us from viewing and analyzing them (all of them were assigned before 2023). For such CVEs and depending on the assigning authority (called CNA), authors usually have to follow up with the CNA to unblind them once the vulnerabilities are publicly disclosed. %
Our analysis found \numcvecategoryinvalidwperc of invalid CVEs, including both CVEs that were formally disputed or rejected as duplicates by the assigning CNA, such as \textsc{Mitre}, and such CVEs where the maintainer of the project considered the report to be invalid or not a bug. In one case, the CVE ID specified in the paper did not match the target, leading us to believe the authors mistakenly reported the wrong number. Three CVEs were claimed by more than one paper, raising questions about who identified and reported them initially. 
A larger number, \numcvecategoryignoredwperc of the CVEs, have been ignored by the maintainers of the respective projects. Investigating this, we found that in \numcveignoredprojectdead cases, the projects were abandoned several years before the bug was found, or the projects had not found widespread adoption (with a single digit number of stars and forks on GitHub). In these cases, the perceived need to report many vulnerabilities in a paper appears to be the driving factor in requesting a CVE for such bugs. 

Studying why some bug reports were ignored while other bugs were fixed, we found that maintainers tend to ignore issues such as memory leaks in client-side software, for example, an assembler. The reasoning appears to be that the program does not run continuously and is not exposed to external attackers. 
Many of the ignored CVEs were segmentation faults in \texttt{mjs} or \texttt{yasm}. The bug tracker of \texttt{mjs} appears to be flooded with similar fuzzer-generated bug reports, while the project has not received an update for two years. Similarly, the maintainer of \texttt{yasm} has moved to other projects, only occasionally merging pull requests. As security researchers usually only drop the bug details without proposing a patch, these issues remain unfixed.
While studying papers, we noticed that several papers claim a specific number of CVEs credited to their work but do not specify any identifier, making it difficult to track them. Interestingly, \numpapersallgoodcves of the \numpaperscveanalysis papers report only CVEs that all have been fixed, accounting for \numpappersallgoodcvescvesum of the CVEs. 

In summary, the need to show a fuzzer's real-world impact results in a large number of unwarranted CVEs, leading to a situation where only \numcvefixedfixedwperc of the \numcvesanalyzed assigned CVEs are valid and have been fixed, while many are what one maintainer referred to as ``fuzzer fake CVEs''~\cite{fuzzerfakecve}. Creating such invalid vulnerabilities causes multiple problems: It unnecessarily alerts people, reduces maintainer acceptance of fuzzer findings, and raises the expectations for subsequent papers to find a similar number of vulnerabilities.

\begin{mdframed}[style=insightstyle]
\numcvecategoryignoredperconly of the CVEs have been ignored and remain unfixed, \numcvecategoryinvalidperconly are invalid. \numcvecategoryblindedperconly are reserved, eluding analysis.
\end{mdframed}

\subsubsection{Statistical Evaluation}%
\label{sec:statistical_evaluation}

To confirm the results obtained in the evaluation, a statistical evaluation is highly recommended~\cite{klees2018evaluating, paassen2021senf} to detect whether the observed difference is significant or by chance. In practice, the most common approach is to compare the final coverage values achieved by different fuzzers across multiple runs.

In general, a frequently used test for the comparison of the means of two sample sets---such as the coverage values of two fuzzers operating on the same target---is the t-test. %
While powerful for the detection of differences, it requires strong assumptions. In particular, the samples have to be approximately normally distributed. This is particularly true for small sample sizes, such as $n \approx 10$. 
To avoid these strong assumptions, the Mann-Whitney or the similar U-test (called Mann-Whitney U-test to emphasize their equivalence subsequently~\cite{Sachs.1984}) is often used. Here, the two samples are assumed to have the same unknown distribution except for a potential shift. The test statistics for the Mann-Whitney U-test is mainly based on the sum of ranks of the two samples in the joint sample. This results in a test for the difference of distribution medians, which is rather robust w.r.t.\ assumptions that do not hold. For a more detailed discussion of such tests, we refer to Sachs' work~\cite{Sachs.1984}.

However, the Mann-Whitney U test can have low power, especially for small sample sizes. Suppose, for example, that we have two samples of three runs that achieved the following coverage:
\[x=(1000,1002,1001),\quad y=(1208,1207,1205)\]%
As is easy to see, these samples are strongly separated, and it is hard to explain these results assuming the similarity of the samples' distributions. 
Yet, the Mann-Whitney U test will not reject the hypothesis of no difference for a significance level $\alpha=5\%$. Even worse, it will never reject samples of this size, since it only uses the ordering of the observations, and the probability of two samples of size $3$ generated from the same distribution to show this pattern of full separation on the real line has a probability $>5\%$. In other words, we cannot use the Mann-Whitney U test to statistically confirm that the difference between two fuzzers is significant if only three trials have been conducted. 
Such situations frequently arise if sample sizes are small or observations cannot be approximately described by a parametric distribution that depends only on few parameters, such as a normal distribution. 

\begin{figure}[bt]
   \centering
   \graphicspath{{graphics}}
   \def\svgwidth{0.99\columnwidth}
   \begin{scriptsize}
        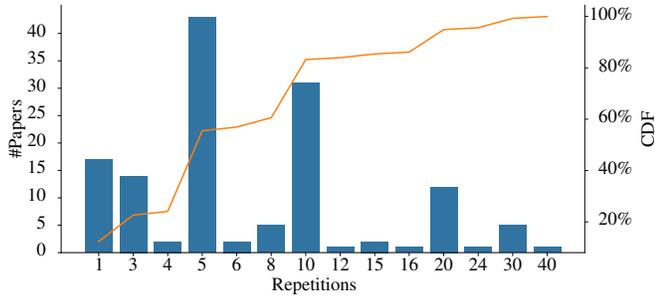
   \end{scriptsize}
   \vspace{-1em}
   \caption{Distribution of trials used in practice and cumulative distribution function (CDF). \numpapersmultipletrials papers use a different number of trials for different experiments; we include all numbers in this case. Further \numtrialsnone papers fail to specify the number of trials.}%
   \label{fig:trial_distribution}
\end{figure}

In summary, a statistical evaluation should use a sufficient number of trials, ideally 10 or more, and use a robust test. 
Studying the trials used in the \numpapers analyzed papers, we find that 1, 3, 5, 10, or 20 trials are the most common repetitions chosen. Figure~\ref{fig:trial_distribution} provides a detailed distribution. Overall, \numtrialslessthantenwperc of the papers use fewer than 10 trials in at least one experiment (\numpapersmultipletrials papers use a different number of trials throughout their paper). Even worse, \nummwunostatisticaltestwperc conduct no statistical test at all. Only \nummwumannwhitneyuwperc of the papers run a Mann-Whitney U test to measure statistical significance, which---paired with few trials---risks that it may never reject the hypothesis. We find that \nummwufewtrialsfiveorlesswperc of the analyzed papers conduct a Mann-Whitney U-test while having five or less trials. One work reports p-values without specifying how they have been derived. Interestingly, we found no other tests, such as bootstrap-based ones, being used, despite being recommended by Klees~\etal~\cite{klees2018evaluating}. Beyond measuring statistical significance, it is recommended to quantify the \emph{effect size}, for example, using Vargha and Delaney's $\hat{A}_{12}$ test~\cite{vargha2000critique}. Yet, we find that only \numeffectsizevarghaanddelaneystestwperc of studies conduct this test; \numeffectsizeotherwperc rely on other means to specify the effect size, leaving us with \numeffectsizenonewperc not using any test to measure the effect size. 

Beyond the use of statistical tests, we find that \numuncertaintynomeasureofuncertaintywperc of the papers provide no measure of \emph{uncertainty}, for example, %
intervals in coverage plots or the standard deviation. This makes it difficult to assess the robustness of reported results, especially considering the inherent randomness in fuzzing runs.

\begin{mdframed}[style=insightstyle]
\nummwunostatisticaltestperconly of the works use no statistical test to assess their results, and \nummwufewtrialsfiveorlessperconly use too few trials to achieve robust outcomes. \numuncertaintynomeasureofuncertaintyperconly provide no measure of uncertainty.
\end{mdframed}

\subsubsection{Threats to Validity}
Scientific works often use a dedicated section on \emph{threats to validity} to enumerate, reflect, and address any issue that could potentially render their evaluation invalid. However, when studying how many of the \numpapers analyzed papers provide such a section, we find that only a minority of \numthreatstovaliditywperc of the papers does so.

\section{Artifact Evaluation}\label{sec:evaluation}

Beyond studying the evaluation outlined and described in the papers, we select \numcasestudies papers and study their artifacts. This allows us to assess the practical reproducibility of fuzzing research and provide recommendations grounded in practice. 
As selection criteria, we pick four recent papers from 2023 and focus on security venues featuring an artifact evaluation. In our experience, papers undergoing an artifact evaluation process provide enhanced documentation and significantly ease the process of setting up a particular tool. However, we test papers that have not undergone artifact evaluation as well to gain a more complete picture. Note that all papers we chose as case studies had attracted our attention during the initial reading for the literature survey in terms of evaluation setup or execution.

In the following, we discuss our lessons learned, pitfalls, and how fuzzing artifacts can be further improved to enhance their reproducibility. 
Again, we emphasize that it is not our intention to point fingers at specific works but rather to highlight potential pitfalls that researchers in this area should be aware of.
More information on all case studies is available in dedicated reproduction repositories on GitHub: \url{https://github.com/fuzz-evaluator/}. %
Despite our best efforts, our reproduction may contain errors. If we become aware of any, we will update the respective reproduction repositories on GitHub. 

\paragraph{\bfseries Author Contact} We have anonymously contacted the authors of all case studies and brought up our findings for discussion with them, asking for their help, confirmation, or clarification.
Five groups have responded to our mails. Where desired by the authors, we publish a statement of them alongside our reproduction artifact. %

\paragraph{\bfseries Setup} All our experiments were performed on two servers running Ubuntu 22.04 with 196~GB RAM, one with an Intel Xeon Gold 6230R CPU with 52 cores at 2.10GHz, and the other with an Intel Xeon Gold 6230 CPU with 40 cores at 2.10GHz (for consistency, a case study was fully run on one type of server or the other). We use the settings provided by the original papers where sensible, otherwise we run 10 trials for 24 hours each, restricting each fuzzer to a single core.

\subsection{Case Study: Artificial Runtime Environment and Unique Crashes}

Our first case study is MemLock~\cite{wen2020memlock}, published at ICSE'20, which proposes to use memory usage as additional feedback. This way, the paper aims to identify resource exhaustion bugs, such as stack exhaustion.

\paragraph{\bfseries Artifact status} MemLock has undergone artifact evaluation and received the \textit{available} and \textit{reusable} badges. Our additional experiments can be found at \url{https://github.com/fuzz-evaluator/MemLock-Fuzz-eval}.

\paragraph{\bfseries Observations} After studying the paper and artifact, we observe the following:
\begin{enumerate}
    \item According to the artifact but not documented in the paper, the authors artificially alter the runtime environment of one target and lower the maximum stack size. Manually limiting the stack size makes it easier to trigger stack overflow bugs, one of the declared goals of the presented technique.
    \item MemLock, similar to many other fuzzing papers, relies on unique crashes as reported by AFL to draw conclusions on the fuzzer's performance. This metric is generally unreliable since a unique crash depends on the set of exercised edges; it does not reflect the number of actual bugs. Here, MemLock's use of the call stack depth as additional feedback may lead to an inflated number of ``unique'' crashes per root cause.
    \item To demonstrate practical impact, MemLock reports 26 CVEs. We found multiple cases among them where up to five CVEs were requested and assigned for a single bug report, to which none of the maintainers responded.
    \item MemLock's artifact is based on PerfFuzz~\cite{lemieux2018perffuzz} (itself an AFL-derivative), but the paper suggests it is based on AFL\@.
\end{enumerate}

We design three experiments to analyze and reproduce MemLock's performance. For full details, we refer the interested reader to our reproduction artifact.

\paragraph{\bfseries Experiment 1: Artificial Runtime Limits} We first study the impact of artificially lowering the stack size for the target \texttt{flex}, which was not documented in the paper. After recreating the setup and running the fuzzing campaign with and without the artificial limit, we observe that MemLock finds the claimed crashes only with the artificially lowered limit. While memory corruption bugs may warrant discussing artificial scenarios, we believe memory exhaustion created through artificial limits cannot be considered realistic. In any case, we recommend documenting such limits in the paper.

\paragraph{\bfseries Experiment 2: Unique Crashes} 
We investigate whether superiority claimed due to \emph{unique crashes} persists when examining the underlying bugs and root causes. 
Using a developer patch and manual triaging, we identify the underlying bugs for three evaluation targets and find that \afl finds four bugs, while MemLock locates only three, even though it finds significantly more unique crashes.

\paragraph{\bfseries Experiment 3: Reported CVEs} When studying the reported vulnerabilities, we noticed that six CVEs, CVE-2020-36370 to CVE-2020-36375, refer to the same bug in \texttt{mjs}. This bug was never acknowledged by the maintainers of \texttt{mjs}. This pattern repeats for other groups of CVEs.

\begin{mdframed}[style=insightstyle]
\textbf{Lessons learned:}
Unique crashes are not a reliable metric; instead, we suggest using (known) bugs. We recommend not using artificial runtime environments without good reason and, if done, documenting such limits. We strongly recommend against the practice of obtaining as many CVEs as possible. Real-world impact should not be measured based on the number of assigned CVEs.
\end{mdframed}

\subsection{Case Study: Exaggerated Vulnerabilities}

For the next case study, we selected SoFi~\cite{he2021sofi}, published at ACM CCS'21. This work aims to use a reflection-based analysis to create a syntactically and semantically valid but diverse set of seeds for fuzzing JavaScript engines.

\paragraph{\bfseries Artifact status} Artifact evaluation was not available for SoFi, but the authors released the source code via an independent web page~\cite{sofiartifact}.
While trying to set up the artifact, we noticed that crucial parts of the source code were missing. The authors stated they would release the missing pieces once the code is polished~\cite{sofiartifact}, but did not react to our e-mails asking for access to the code. Without a chance to reproduce the artifact, we solely studied the paper, in particular the reported vulnerabilities summarized in Table~2 of their paper, entitled ``Summary of discovered vulnerabilities''~\cite{he2021sofi}.

\paragraph{\bfseries Observations} We find that all seven vulnerabilities claimed in the actively used modern browser engines (\ie v8, SpiderMonkey, and JavaScriptCore) are invalid and have been rejected by the respective developers, six out of seven even \emph{before} the conference submission deadline. While SoFi manages to find confirmed vulnerabilities in other programs, we believe it is important to not oversell results by claiming to have found vulnerabilities in browser engines, when in fact they were not a bug at all. We assume that the bug report IDs were blinded, as is common practice for submission, such that the reviewers could not verify the validity of the presumed vulnerabilities.

\begin{mdframed}[style=insightstyle]
\textbf{Lessons learned:}
We highly discourage marketing invalid bug reports as a vulnerability. Feedback from the developers must be taken into account (especially if bug reports are rejected by the developers). Pledges to release the source code should be kept.
\end{mdframed}

\subsection{Case Study: Missing Baselines}

DARWIN~\cite{jauernig2023darwin} was published at NDSS'23 and honored with a \emph{Distinguished Paper Award}. The paper focuses on improving mutation scheduling. More specifically, the authors propose to use an evolution strategy and dynamically adapt the mutation selection to the target under test. 

\paragraph{\bfseries Artifact status} Artifact evaluation was not available to DARWIN, but the authors publicly released an artifact. Our reproduction artifact is available at \url{https://github.com/fuzz-evaluator/DARWIN-eval}.

\paragraph{\bfseries Observations}
Analyzing the paper and artifact, we found a number of issues:
\begin{enumerate}
    \item Coverage differences between DARWIN and tested baselines on \fuzzbench are not statistically significant nor consistent with the paper's \fuzzbench results.
    \item The results on MOpt~\cite{lyu2019mopt} listed in the DARWIN paper indicate that the port implemented for MOpt may have erroneously restricted the number of usable mutations. We find that this strongly influences the results.
    \item The artifact appears to be based on Git tag 2.55b of Google's AFL fork and not 2.54b, as listed in the paper.
    \item The artifact does not provide the AFL~2.55b port for MOpt or their baseline AFL-S, preventing reproduction or analysis.
\end{enumerate}

We design three experiments to analyze DARWIN\@. More experiments and details are available in our artifact.

\paragraph{\bfseries Experiment 1: Coverage} 
We use \fuzzbench to reproduce DARWIN's coverage measurements (in particular, Table III of their paper). Running all targets for 24 hours, we compare it against AFL~2.55b and MOpt, which is based on AFL~2.52b. Notably, we do not use DARWIN as configured in \fuzzbench but follow the author's recommended configuration (see Experiment~3). In our \fuzzbench results, MOpt does not show the major performance degradation shown in the paper results. Overall, \fuzzbench ranks DARWIN above MOpt and AFL, both by score and rank. In individual targets, DARWIN is the best performer in nine of the targets, but only with statistical significance in four. 
Our results show the difference between DARWIN and its baselines to be less than reported in Table III of their paper.
Where they find DARWIN's median relative coverage to be the highest for 15 out of 19 targets, we find this to be the case for 4 out of 18 targets\footnote{\fuzzbench has meanwhile removed the target \texttt{php}.} (DARWIN is worse than at least one baseline in two cases and tied with at least one baseline in the other cases).
Note that the original paper evaluates over a six hour period instead of the 24 hours recommended by Klees~\etal~\cite{klees2018evaluating}. While we provide the statistical data for the 24 hour data here, we emphasize that the results reported in the paper for the six hour mark are similarly not reproducible and invite the reader to view our full evaluation report data available on GitHub.

In summary, our results show a similar tendency to their paper, but the difference observed between DARWIN and its baselines is smaller.
Notably, DARWIN reports a coverage improvement of only 1.73\% over AFL, making it difficult to judge the difference between these fuzzers meaningfully.

\paragraph{\bfseries Experiment 2: New Baseline} 
We propose a second baseline to test DARWIN's contribution of a dynamically adapting mutation selection: we replaced its proposed weighting with a random selection (that is reweighted at a constant interval).  
This implementation, DARWIN$_{RAND}$, provides a new baseline that allows to better judge DARWIN's contribution, as any improvement can be directly attributed to DARWIN's evolutionary algorithm rather than other fuzzer implementation details, such as dynamically adapting mutation selection.
We find in our \fuzzbench results no statistical significant difference between DARWIN and DARWIN$_{RAND}$, meaning we were unable to demonstrate that the evolutionary aspects of DARWIN's approach significantly contributed to the improvement compared to randomly changing mutation selection over time.

\paragraph{\bfseries Experiment 3: Per-Seed Mutation Scheduling} 
After contacting the authors, they noted that the per-seed mutation scheduling (\texttt{-p} flag) set by \fuzzbench should be disabled for the evaluation because it worsens performance and was not intended as part of the paper. To confirm this, we separately evaluated DARWIN with and without per-seed mutation scheduling on seven targets: we found that disabling the per-seed mutations slightly improved performance overall, leading to higher median coverage in some targets,
but not statistically significantly so for any target by Mann-Whitney U. We have used the author-recommended configuration (no \texttt{-p} flag) for Experiments~1 and~2.

\begin{mdframed}[style=insightstyle]
\textbf{Lessons learned:}
A baseline suited to test the proposed technique is necessary to detect differences that can be attributed to the proposed technique rather than the new fuzzer implementation as a whole.
We further recommend publishing all evaluation artifacts, also including benchmarking reports and raw data.
\end{mdframed}

\begin{table}[b]
    \centering
    \caption{Comparing the code coverage reported by FuzzJIT to our measurements.}%
    \label{tab:fuzzjitCoverage}
    \begin{tabular}{l|ccc|c}
        \toprule
        & \multicolumn{3}{c|}{\textbf{Reported}} & \textbf{Measured}\\
        \textbf{Engine} & \textbf{Fuzzilli} & \textbf{FuzzJIT} & \textbf{Rel. Increase} & \textbf{Rel. Increase} \\
        \midrule
        JSC & 16.47\% & 21.90\% & 33\% & ~-2\% \\
        V8  & 13.82\% & 16.67\% & 21\% & ~-3\% \\
        SM  & 15.53\% & 17.97\% & 16\% & -12\% \\
        \bottomrule
    \end{tabular}
    \vspace{-1em}
\end{table}

\begin{table}[tb]
    \centering
    \caption{Comparing the semantic correctness rate reported by FuzzJIT to our measurements.}%
    \label{tab:fuzzjitSematicCorrectness}
    \begin{tabular}{l|rr|rr}
        \toprule
        & \multicolumn{2}{c|}{\textbf{FuzzJIT}} & \multicolumn{2}{c}{\textbf{Fuzzilli}} \\
        \textbf{Engine} & \textbf{Reported} & \textbf{Measured} & \textbf{Reported} & \textbf{Measured}  \\
        \midrule
        JSC & 90.33\% & 65.88\% & 62.80\% & 66.56\% \\
        V8  & 97.04\% & 63.67\% & 64.34\% & 66.74\% \\
        SM  & 93.28\% & 63.93\% & 64.13\% & 67.47\% \\
        \bottomrule
    \end{tabular}
\end{table}

\subsection{Case Study: Non-reproducible Measurements}

A recent paper published at USENIX'23, FuzzJIT~\cite{wang2023fuzzjit}, aims to detect bugs in JIT compilers, including those used in modern browsers.

\paragraph{\bfseries Artifact Status} 
FuzzJIT underwent artifact evaluation and was awarded the \emph{available} and \emph{functional} badges. Our reproduction artifact can be found at: \url{https://github.com/fuzz-evaluator/fuzzjit-eval}.

\paragraph{\bfseries Observations}
After studying the paper and testing the artifact, we observe several shortcomings:

\begin{enumerate}
    \item Coverage does not reproduce as outlined in the paper; in our experiments, FuzzJIT performed worse than Fuzzilli on all targets.
    \item Reported improvements of the semantic correctness rate did not materialize in our experiments.
    \item It is not possible to study the bugs found because the time frame, engine versions, and resources spent were not specified in the paper, hindering fair reproduction.
\end{enumerate}

We design two experiments to analyze the claims of FuzzJIT in more detail. %

\paragraph{\bfseries Experiment 1: Code Coverage} When trying to reproduce code coverage, we find significantly different results. As shown in Table~\ref{tab:fuzzjitCoverage}, FuzzJIT reports a code coverage improvement of up to 33\% over Fuzzilli. In stark contrast, our experiments show a code coverage decrease of -2\% to -12\%. Despite searching for the cause, we find none explaining this difference.
We speculate that the negative outcome of the comparison experiment is a consequence of benchmarking with different versions of Fuzzilli. This is based on the observation that the state-of-the-art fuzzers compared to in the evaluation are taken from UniFuzz~\cite{li2021unifuzz}, which uses an outdated version of Fuzzilli; FuzzJIT itself is based on a more recent version of Fuzzilli. Unfortunately, the authors have not responded to our request for help.

\paragraph{\bfseries Experiment 2: Semantic Correctness Rate} Besides code coverage, FuzzJIT also evaluates the semantic correctness rate of generated samples, \ie the number of samples that do not raise an uncaught exception during execution in the JS engine.
As shown in Table~\ref{tab:fuzzjitSematicCorrectness}, we could \emph{not} measure any improvement of the semantic correctness rate, contrasting the paper's claim of a significant improvement. %

\begin{mdframed}[style=insightstyle]
\textbf{Lessons learned:}
Relying on outdated baseline versions can create a distorted picture of a fuzzer's performance.
Authors should ensure that they use the latest version of all tools for comparison.
\end{mdframed}

\subsection{Case Study: Uncommon Metrics}

Published at USENIX'20, EcoFuzz~\cite{yue2020ecofuzz} proposes to replace AFL's seed scheduling algorithm with a version relying on the adversarial multi-armed bandit model. This way, EcoFuzz finds more paths while generating less seeds. 
\paragraph{\bfseries Artifact status} EcoFuzz has undergone artifact evaluation and was awarded the \emph{passed} badge, indicating that the artifact is available and ready to be reproduced. Our independent reproduction repository is located online at \url{https://github.com/fuzz-evaluator/EcoFuzz-eval}. 

\paragraph{\bfseries Observations} When studying the paper and artifact, we noticed that the evaluation deviates from typical fuzzing evaluations: The work does not report achieved code coverage over time. 
Instead, the paper visualizes the total number of paths discovered over executions. This aligns with the paper's goal of finding more path (bandits in EcoFuzz's multi-armed bandit model) with fewer executions (trials in the model). 
The presented results may lead readers to infer that a higher number of total paths equates to higher code coverage, which is not necessarily true.

\begin{figure}[tb]
   \centering
   \graphicspath{{graphics}}
   \def\svgwidth{\columnwidth}
   \begin{tiny}
       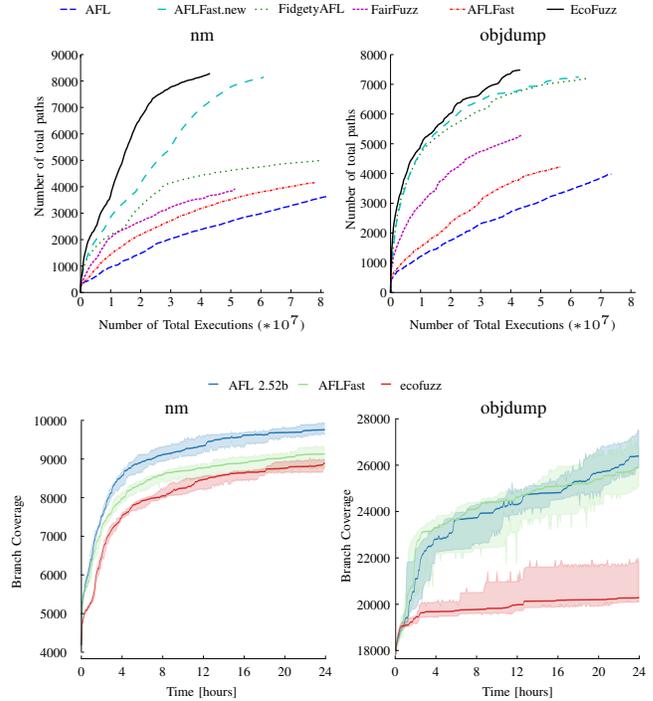
   \end{tiny}
   \vspace{-1.5em}
   \caption{The upper two graphs published in the EcoFuzz paper~\cite{yue2020ecofuzz} show a strong advantage over all competitors on the non-standard metric \textit{number of totals paths over the number of total executions}. The two plots at the bottom compare EcoFuzz on the standard metric \textit{branch coverage over time}. On the standard metric, EcoFuzz performs significantly worse.}%
   \label{fig:ecofuzz_cov}
\end{figure}

\paragraph{\bfseries Experiment: Code Coverage} We design an experiment in \fuzzbench where we compare EcoFuzz against its best-performing competitor, AFLFast, and its baseline, AFL\@. We test these fuzzers on three targets, \texttt{nm}, \texttt{libpng}, and \texttt{objdump}, where the original evaluation\footnote{The evaluation used \texttt{readpng}, which internally uses \texttt{libpng}, while we use \texttt{libpng\_read\_fuzzer} as bundled with \fuzzbench.} found EcoFuzz to be the fuzzer to find the \emph{most} paths. 
Our results, shown in Figure~\ref{fig:ecofuzz_cov}, demonstrate that EcoFuzz achieves \emph{less} code coverage than the other fuzzers in all scenarios, except for a statistically insignificant one, where it performs similar to AFLFast on \texttt{libpng}. This underlines that finding more paths does not necessarily translate to achieving a higher coverage.  
The full results and the generated \fuzzbench reports can be found in our reproduction repository.

Corresponding with the authors, they state they have been following fuzzing evaluations at the time that focused on path coverage, and they have confirmed that EcoFuzz may cover fewer branches on some binaries, stating that its goal is to optimize for paths over executions rather than branches over time. %

\begin{mdframed}[style=insightstyle]
\textbf{Lessons learned:}
A fuzzer may excel at one metric but not on another; hence, selecting a suitable set of evaluation metrics is crucial to provide a reader with the full picture. Evaluating on established metrics is required, as new metrics may imply a completely different picture.
\end{mdframed}

\subsection{Case Study: Unclear Documentation}

Another paper published at USENIX'23, Polyfuzz~\cite{li2023polyfuzz}, targets programs containing code in different languages, such as interpreter languages calling into native bindings. 

\paragraph{\bfseries Artifact status} 
PolyFuzz has been awarded the \emph{available} badge. %
Our reproduction artifact is available at \url{https://github.com/fuzz-evaluator/PolyFuzz-eval}.

\paragraph{\bfseries Observations} While studying the artifact, we noticed irregularities regarding the seed sets used by PolyFuzz compared to the other fuzzers. An example of such a case is the \verb|image_load| harness for the Python image processing library Pillow. In this particular case, the fuzzer Atheris gets 39 seed files, while PolyFuzz's seed directory has 58 files. 

\paragraph{\bfseries Experiment: Fair seed allocation} We intended to run both fuzzers with their respective seed sets to measure the impact of these different seed sets on the coverage.
Unfortunately, the authors' extension of Atheris (called Atheris-Cext in the PolyFuzz paper), which would allow to compute combined coverage for both Python and the native code, was not released alongside their artifact.
Hence, as proxy measurement, we compute the initial coverage achieved by PolyFuzz on both seed sets. 
For the seed set given to Atheris, PolyFuzz covers 218 edges, while for its own seed set, it covers 814 edges. Evidently, one seed set provides more than three times as much coverage as the other, giving PolyFuzz a headstart during the evaluation.

When contacted, the authors clarified that they did not keep the seed sets from their evaluation, but they assured us that they used the seeds from the corresponding benchmarks for all fuzzers.

\begin{mdframed}[style=insightstyle]
\textbf{Lessons learned:}
Seeds have an impact on fuzzer performance. We recommend to give all fuzzers the same set of seeds and to publish the seeds used.
\end{mdframed}

\subsection{Case Study: Incomplete Artifact}
\firmafl~\cite{zheng2019firm-afl}, published at USENIX Security'19, aims to fuzz Linux-based IoT firmware via augmented process emulation. To do so, the core fuzzing loop targets a single binary under user-mode emulation, while selectively forwarding system calls to a full-system emulator.

\paragraph{\bfseries Artifact status} Artifact evaluation was not available to \firmafl, but different versions of its source code are publicly available across multiple repositories.
Our reproduction artifact is available at \url{https://github.com/fuzz-evaluator/firmafl-eval/}.

\paragraph{\bfseries Observations}
During our analysis of the artifact, we noticed that the repository lacks documentation. Crucial steps are missing, like correct build instructions for different configurations, making it hard for researchers to reuse the artifact and set up the fuzzer and its environment correctly. Furthermore, when setting up the experiments, we noticed that some of the experiment configuration files were missing and target harnessing is heavily inlined with core emulation logic. Not only do these issues hinder extensibility, but they also prevented us from getting all targets working to reproduce the \firmafl experiments.
The fuzzer binaries are shipped in a pre-compiled binary version and fail to build from the provided source code. Moreover, the provided baseline uses an older version of AFL (2.06b), while the augmented mode uses AFL v2.52b.

\paragraph{\bfseries Experiment: Crash Triggers} Being the only experiment with enough documentation to reproduce, we aim to measure the number of crashes produced by both the augmented and full-system emulator versions. 
We were able to run fuzzing campaigns for 9 out of 11 targets, where one of them only ran for the baseline and not \firmafl. The remaining two targets lack the required target-specific configurations. Unfortunately, we could only partially reproduce the claims as presented in the \firmafl paper and observed one case where the baseline performed better than \firmafl.
The full results of our experiments can be found in our reproduction repository.

\begin{mdframed}[style=insightstyle]
\textbf{Lessons learned:}
While it is unreasonable to expect each academic artifact to be of production quality, we recommend to strive for a reasonable level of readability and documentation that allows others to understand and use the code, thus promoting reproducibility.
\end{mdframed}

\subsection{Case Study: Unfair Coverage Measurements}
The final case study analyzes FishFuzz~\cite{zheng2023fishfuzz}, published at USENIX'23. The paper proposes an input prioritization strategy based on a multi-distance metric that allows for optimizing the fuzzing efforts towards thousands of targets (\eg sanitizer labels) in the sense of direct fuzzing.

\paragraph{\bfseries Artifact status} %
FishFuzz has received the \emph{available} and \emph{functional} badges. Our additional experiments are available at \url{https://github.com/fuzz-evaluator/FishFuzz-eval}.

\paragraph{\bfseries Observations} When studying the artifact in detail, we notice that FishFuzz's way of measuring coverage may erroneously give FishFuzz an unfair edge.
From all evaluated fuzzers, FishFuzz was the only fuzzer to place coverage instrumentation not only within the actual target but also in the added ASAN instrumentation. Consequently, FishFuzz also stored inputs that exercised new coverage in the instrumentation; other fuzzers discarded these inputs, as no new coverage was observed.
This became a problem when the binary instrumented by FishFuzz was used for coverage measurements for all fuzzers during evaluation since---by design---only FishFuzz would keep inputs exercising coverage in the ASAN instrumentation.

\begin{figure}[tb]
    \centering
    \includegraphics[width=0.82\columnwidth]{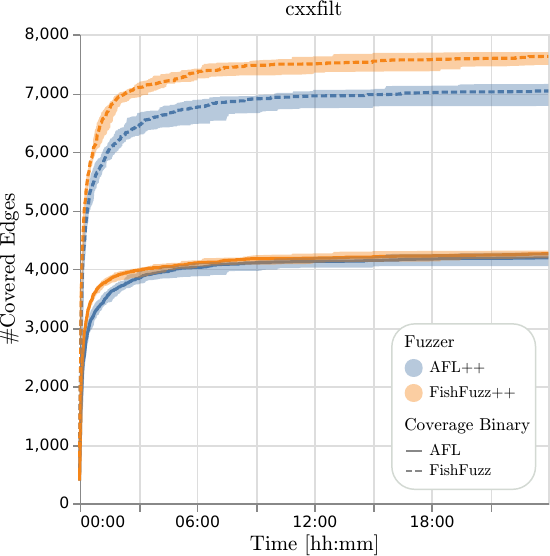}
    \caption{Median coverage over time for \texttt{cxxfilt}: In one case, we measure coverage via a standard AFL binary and, in the other we use FishFuzz's binary that contains additional coverage instrumentation. For each fuzzer, the target was run 10 times for 24h each. The displayed intervals enclose all ten runs of the respective fuzzer. If the coverage is measured on the biased binary with FishFuzz instrumentation~(\inlineincludegraphics{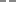}), FishFuzz++ finds on average 8.44\% more edges than AFL++. Measuring coverage on a standard AFL binary~(\inlineincludegraphics{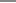}) (without additional instrumentation introduced by FishFuzz), the coverage delta is only 1.69\%.}%
    \label{fig:fish_fuzz_cxxfilt_coverage}
    \vspace{-1em}
\end{figure}

\paragraph{\bfseries Experiment: Fair coverage measurement} 
To demonstrate the impact of measuring coverage in instrumentation code, we measure the coverage for a binary both with and without FishFuzz instrumentation.
The result %
is depicted in Figure~\ref{fig:fish_fuzz_cxxfilt_coverage}. If the FishFuzz coverage binary is used for coverage computation, FishFuzz covers 8.44\% more edges on average over all runs. When using a binary with standard \afl instrumentation (\ie where coverage is not measured in the additional instrumentation), the observed coverage increase is reduced to 1.69\%. Furthermore, the total number of edges is considerably smaller, showing that edge counts between different binaries do not translate. Note that both coverage binaries rely on colliding bitmaps since the artifact tooling of FishFuzz expects standard AFL bitmaps to be used. We recommend to not use colliding bitmaps for coverage measurements.

\begin{mdframed}[style=insightstyle]
\textbf{Lessons learned:}
Unintended side effects may skew coverage measurements; we recommend using standardized methods of measuring coverage.
\end{mdframed}

\section{Revised Best Practices for Evaluation}%
\label{sec:best_practices}

Based on our literature analysis and the case studies, we now provide recommendations on ensuring a fair and reproducible fuzzing evaluation. 
A comprehensive checklist that summarizes these recommendations is available in our GitHub repository at \url{https://github.com/fuzz-evaluator/guidelines}.
Overall, we recommend that authors thoroughly review the \emph{threats to validity} for their respective works to reflect potential issues that could invalidate their evaluation.

\subsection{Reproducible Artifact}
For reproducibility, it is crucial to open-source the source code including documentation. We highly recommend participating in an artifact evaluation if available. Furthermore, it is essential to
\begin{inparaenum}[(i)]
\item specify the exact versions of targets (and harnesses) and fuzzers used for comparison,
\item use runtime environment abstractions, such as Docker (where feasible),
\item name the baseline on which the new technique is implemented upon (if any) as well as its version, and avoid squashing commits of this baseline.
\end{inparaenum}
In the long term, a mandatory artifact evaluation as part of the submission process could improve the quality and reproducibility of research artifacts.

\subsection{Targets under Test}
Selected evaluation targets should form a representative set that shows strengths of the proposed approach and allows for comparability with previous work. It is therefore desirable to include targets that have been tested in other works.
Actions such as patches applied to targets should be explained. If a fuzzer has certain restrictions (such as symbolic execution-based techniques not being able of modeling all syscalls), we recommend outlining those. We also highly recommend using well-established benchmarks, such as \fuzzbench, to facilitate easy reproducibility.

\subsection{Comparison to Other Fuzzers}
It is crucial to compare against the state of the art in the respective field and the baseline (if any) on which the new technique is implemented.
This also includes well-established and actively maintained fuzzers, such as AFL++. Including the new fuzzer in benchmarks such as \fuzzbench allows for comparing against a wide range of fuzzers.
If presenting a new technique with separable design choices, review them individually via ablation studies, for example, by designing baselines that successively enable or disable individual components.

\subsection{Evaluation Setup}
The chosen evaluation setup should be well documented. This entails details regarding the used hardware, experiment runtime, number of allocated cores, and processes per fuzzer. The conducted experiments and how to reproduce them should be explained.

For the runtime, we recommend choosing at least 24 hours. Longer runtimes may be appropriate if the evaluated fuzzers do not flatline at the end of the experiment. Regarding CPU cores, choosing a single core may not be representative of modern systems. Special care must be taken to avoid congestion in the kernel when running multiple fuzzers in parallel on one system; even if using Docker, the kernel may become a bottleneck in resolving certain syscalls, unfairly slowing down one fuzzing process. Individual fuzzer instances can be encapsulated in separate virtual machine instances to avoid such situations.

Regarding seeds, we recommend running with uninformed seeds or multiple seed sets. Seeds must be described and accessible (in the case of informed seeds) to allow for reproducibility. All fuzzers should have fair access to all seeds. If using informed seeds, we recommend plotting or analyzing the coverage achieved by the initial seed set. This avoids attributing a high coverage achieved to fuzzer performance instead of the initial seeds.

\subsection{Evaluation Metrics}
A fuzzer comparison should use standardized, well-established metrics (at least as a complementary metric if a technique requires the introduction of a new metric); this includes both coverage and found bugs. Optimally, both code coverage and bug-finding capability are evaluated, as both suffer from individual drawbacks~\cite{klees2018evaluating,boehme2022_covreliability,zeller2019blog}. We recommend using modern benchmarks that aid in setting up the experiment and ensure a fair, bias-free execution. 

It is necessary to specify details such as how coverage is collected, for example, whether it is measured on a non-instrumented binary, translated blocks from an emulator, or using established means such as \texttt{lcov}. Ideally, coverage is not measured using bitmaps with collisions, but using a collision-free encoding or other means. Additionally, the evaluation must ensure that the same notion of coverage is used for each of the compared fuzzers. %

When searching for bugs in new targets to show real-world impact, it is crucial to select reasonable targets, \ie projects that are not insecure by design, have been inactive for years, or are unsuitable for other reasons. We also recommend running other state-of-the-art fuzzers to see whether they find the bugs as well, thereby addressing concerns regarding fuzzing previously untested software. 
Crashes identified by the fuzzer should be deduplicated before opening a report, and the triaging process should be clearly described. When testing crashes, we recommend reproducing them on a binary without fuzzer or coverage instrumentation to avoid reproducibility issues.

Ideally, only maintainers should request CVEs. If they do not request one, researchers can still link to the bug report instead. Requesting multiple CVEs for a single bug or requesting CVEs without coordinating or informing the maintainers must be avoided.
If possible, reporting bugs or CVEs anonymously allows for providing the reviewers with access during submission, such that they can inspect the CVEs or bug reports and assess their validity (as opposed to the current practice of blinding CVEs and bug reports during submission, preventing any analysis by reviewers). That said, we do not believe that having CVEs should be required to show the practical impact of a fuzzer. %

\subsection{Statistical Evaluation}
Any evaluation should be backed by statistical tests. To enable these tests, we recommend running at least ten trials. 
Alternatively, the number of trials can be calculated via an a-priori power analysis to ensure a sufficient sample size leading to statistically significant results~\cite{cohen2013statistical}. This is particularly important if the fuzzer under consideration only slightly outperforms the state of the art, where $n\gg10$ may be required.
To avoid the problems mentioned in Section~\ref{sec:statistical_evaluation}, we recommend an alternative to the widely used Mann-Whitney-U test; permutation tests or resampling tests such as bootstrap methods. These methods avoid strong assumptions regarding a normal distribution.

If more than two fuzzers have been compared for a target, the (bootstrap-based) two-sample t-test is not a good choice, since we would have to perform more than one pairwise comparison to test the null hypotheses of no difference between any of the expected means for the fuzzing methods. This results in the \emph{multiple testing problem}, which is the observation that the probability of at least one false positive result in the set of comparisons performed for a target exceeds the single test level $\alpha$ substantially. The same argument holds for other strategies based on two-sample comparisons such as the Mann-Whitney-U test~\cite{arcuri2011practical}. 

A solution to this problem is the bootstrap version of the \emph{ANOVA method}. If the ANOVA rejects the null hypothesis, it shows at level $\alpha$ that there is at least one pair of fuzzing methods that perform significantly different for the target considered. In a second step, a so-called \emph{Posthoc}-test is performed to determine which pairwise comparisons are significant, \emph{given that the ANOVA has already shown that there are significant differences at all}. Possible \emph{Posthoc}-tests are, for example, the Tukey-Kramer method if all pairwise comparisons among all samples are of interest or the Dunnett method if only the comparisons to a reference method, such as the newly developed fuzzer, are of interest~\cite{Sachs.1984}. 
For a bootstrap version of these algorithms, we propose as a simple solution two-sample t-tests with critical values for rejection based on a bootstrap resampling with replacement of the test statistics. 
Here, for each simulation, the maximum value of the test statistics is used for all pairwise comparisons of interest.
We provide more details, algorithms, and scripts implementing examples for these tests in our artifact at \url{https://github.com/fuzz-evaluator/statistics}. Additionally, evaluations should measure \emph{effect size}, \eg using Vargha and Delaney's $\hat{A}_{12}$ test~\cite{vargha2000critique}, and quantify \emph{uncertainty}, for example, by using intervals in plots.

\section{Conclusion}\label{sec:conclusion}

Reproducibility is a cornerstone of science and the basis for research. 
In this work, we have systematically studied how \numpapers fuzzing papers published in the past six years at leading conferences design their evaluation. Furthermore, we have performed an in-depth analysis of the artifacts of \numcasestudies papers and attempted to reproduce their results. Based on the insights gained, we outlined several potential pitfalls and shortcomings threatening the validity of fuzzing evaluations. Ultimately, we provided revised recommendations and best practices to improve future evaluation of fuzzing research. We published a concise set of guidelines at \url{https://github.com/fuzz-evaluator/guidelines} and welcome community contributions.

\section*{Acknowledgment}

We thank our anonymous shepherd and reviewers for their valuable feedback. Further, we thank Dominik Maier, Johannes Willbold, Daniel Klischies, Merlin Chlosta, and Marcel Böhme (in no particular order) for their helpful comments on a draft of this work.
We also thank the countless researchers with whom we have discussed fuzzing research and how to evaluate it, ultimately paving the way for this work.
This work was funded by the European Research Council (ERC) under the consolidator grant RS$^3$ (101045669) 
and the German Federal Ministry of Education and Research under the grants KMU-Fuzz (16KIS1898) and CPSec (16KIS1899).
Additionally, this research was partially supported by the UK Engineering and Physical Sciences Research Council (EPSRC) under grant EP/V000454/1. The results feed into DsbDtech.

\onecolumn
\begin{multicols}{2}

\end{multicols}

\twocolumn

\appendices
\newpage %

\section{Meta-Review}

The following meta-review was prepared by the program committee for the 2024
IEEE Symposium on Security and Privacy (S\&P) as part of the review process as
detailed in the call for papers.

\subsection{Summary}
This SoK submission selects 150 papers from 2018-2023 published in top-tier security and software engineering venues for fuzzing research. It then performs a meta-evaluation of each paper’s evaluation in terms of experimental design and adherence to generally accepted fuzzing guidelines using Klees et al. as a baseline. In addition, eight papers are subject to artifact evaluation. The conclusions are stark: fuzzing papers continue to fall short of known best practices in conducting rigorous fuzzing research. An updated set of guidelines is then presented.

\subsection{Scientific Contributions}
\begin{itemize}
\item Independent Confirmation of Important Results with Limited Prior Research
\item Addresses a Long-Known Issue
\item Provides a Valuable Step Forward in an Established Field
\item Other (Reproducibility Study)
\end{itemize}

\subsection{Reasons for Acceptance}
\begin{enumerate}
\item Fuzzing is an important research area, and understanding whether fuzzing papers hew to best practices intended to maximize the validity and reproducibility of the results is important
\item The paper uses an overall strong review methodology
\item The paper examines a wide range of fuzzing papers over time and across conferences
\item The paper includes an artifact evaluation on a subset of the reviewed fuzzing papers
\item The paper's observations are significant\end{enumerate}

\end{document}